\documentclass{elsart}
\usepackage{amsmath, amssymb,graphicx,deluxetable} 

\usepackage{amssymb}

\newcommand{\be}{\begin{equation}}
\newcommand{\sun}{\odot}
\begin{document}

\begin{frontmatter}



\title{Cosmic and Galactic Neutrino Backgrounds from Thermonuclear 
Sources}


\author{Cristiano Porciani\corauthref{cor1}}
\corauth[cor1]{Corresponding author}
\ead{porciani@phys.ethz.ch}
\address{Institute of Astronomy, Department of Physics, 
ETH H\"onggerberg HPF G3.1, CH-8093, Z\"urich, Switzerland}

\author{Silvia Petroni\corauthref{cor2}}
\address{Dipartimento di Fisica, Universit\`a di Pisa and INFN Sezione di 
Pisa,via Buonarroti 2, I-56127 Pisa, Italy}
\corauth[cor2]{Also at: Dipartimento di Fisica, Universit\`a di Ferrara, via
Paradiso 12, I-44100 Ferrara, Italy}
\author{Giovanni Fiorentini}
\address{Dipartimento di Fisica, Universit\`a di Ferrara and
INFN Sezione di Ferrara, via
Paradiso 12, I-44100 Ferrara, Italy}

\begin{abstract}
We estimate energy spectra and fluxes at the Earth's surface  
of the cosmic and Galactic 
neutrino backgrounds produced by thermonuclear reactions 
in stars.
The extra-galactic component 
is obtained by combining the most recent estimates of the 
cosmic star formation history and the stellar initial mass function
with accurate theoretical predictions of the neutrino yields all over the
thermonuclear lifetime of stars of different masses.
Models of the structure and evolution of the Milky Way are 
used to derive maps of the expected flux generated by Galactic sources
as a function of sky direction.
The predicted neutrino backgrounds depend only slightly
on model parameters. 
In the relevant 50 keV-10 MeV window,
the total flux of cosmic neutrinos 
ranges between 20 and 65 ${\rm cm}^{-2}\,{\rm s}^{-1}$.
Neutrinos reaching the Earth today have been typically
emitted at redshift $z\sim 2$. Their energy 
spectrum peaks at $E\sim 0.1-0.3$ MeV.
The energy and entropy densities of the cosmic background
are negligible with 
respect to the thermal contribution of relic neutrinos originated in the
early universe.
In every sky direction, the cosmic background is outnumbered by
the Galactic one, whose integrated flux amounts to
300-1000 ${\rm cm}^{-2}\,{\rm s}^{-1}$.
The emission from stars in the Galactic disk contributes more than 95
per cent of the signal.
\end{abstract}

\end{frontmatter}


\section{Introduction}
Almost all our current knowledge of the universe derives from the 
detection of electro-magnetic quanta of different energies.
However, the hot dense regions lying at the centre of most astrophysical
sources are completely opaque to photons, and
their physical properties can only be inferred indirectly.
Additional information on astrophysical objects is gathered by studying the 
properties of cosmic rays. 
Yet, these are 
charged particles and the presence of galactic magnetic fields makes
nearly impossible to trace them back to their sources.
Another limitation which applies to observations of both high-energy
photons and charged particles is given by the so called
Greisen-Zatsepin-Kuz'min effect, namely the creation of
electron-positron pairs on the cosmic infrared
radiation background (for photons) and of pions on the
cosmic microwave background (for protons and heavier nuclei).
In other words, the universe is optically thick to photons above $\sim 10$ TeV
and to protons above $\sim 10 $ EeV on a scale of few
tens of Mpc.
Neutrino astronomy, although still in its infancy, might provide a way to 
circumvent all these limitations. 
In fact, 
neutrinos are stable (so that they can cross long distances), weakly 
interacting (so that they are able to penetrate regions which are opaque to 
photons), and electrically neutral (so that their trajectories are not  
affected by magnetic fields). 
It is then reasonable to expect that neutrino telescopes would allow us to 
get spectacular new views of the universe. 

Regrettably, the same reasons which make neutrinos interesting
astrophysical probes make their detection extremely difficult. 
At present, only the Sun and Supernova 1987A have been 
detected through their neutrino emission. 
This is not surprising,
since, as like as in the electro-magnetic visible band, the Sun is
expected to largely outshine any other astronomical source.  Anyway, 
the detection of solar neutrinos has already led to important
advances.  
By cross-checking the experimental results against the predictions of
standard solar models it was realized that only a fraction
(ranging between a third and two thirds depending on the particle
energy) of the expected solar neutrino flux had been detected
(see e.g. \cite{BAH02} and references therein). 
A possible interpretation is that 
part of the solar electron neutrinos 
transform into different particles. 
This indication, demonstrated by the Sudbury Neutrino Observatory (SNO) 
detection of non-electron neutrinos 
from the sun \cite{AHM01} 
and confirmed by the disappearance of reactor antineutrinos
in the Kamioka Liquid Scintillator Anti-Neutrino Detector
(KamLAND) \cite{EAL03}, 
requires modifying the minimal standard model of electro-weak
interactions so as to allow for neutrino flavor mixing and oscillations.

An important problem for the development of neutrino astronomy is the 
determination of sky
backgrounds as a function of energy. Many different
sources might contribute, ranging from distant galaxies to Galactic
supernovae.
In this paper we estimate fluxes and energy spectra at the Earth's surface  
of the cosmic and Galactic neutrino backgrounds generated by
thermonuclear activity in stars.

The nuclear fusion reactions which turn 
hydrogen into helium produce 
two electron neutrinos and release about 27 MeV of energy 
(including the neutrino budget).
The generation of a solar luminosity by hydrogen-core burning then implies
the emission of about $2\times 10^{38}$ neutrinos per second.
In consequence,
we expect the existence of a 
diffuse and nearly isotropic neutrino background originating from all stars 
in the
universe that are and/or have been thermonuclearly active.
At the same time, the fact that we live within the disk of a spiral galaxy
should imprint characteristic features on the angular
distribution of the neutrino flux.

Despite the large effort aimed at estimating the amplitude and spectrum
of the cosmic background generated by 
Type II supernovae \cite{HW97,KSW00,AST03}
and Gamma-ray bursts \cite{NAL03},
the current literature still lacks of accurate
predictions concerning the cosmic and Galactic stellar neutrino backgrounds. 
A first attempt in this direction has been performed by
Hartmann et al. \cite{HAL95}
who, however, 
only considered main-sequence stars and used semi-analytic stellar
models to compute the neutrino emission rates.  Accurate numerical
determinations of the neutrino yields over the whole H and He burning
phases have been presented by Brocato et al. \cite{BAL98}.
These results
have been used by the same authors to estimate the present neutrino
flux at the Earth's surface due to Galactic sources. Their Monte Carlo
models give fluxes ranging from 3 to 24 
${\rm cm}^{-2}\,{\rm s}^{-1}\,{\rm deg}^{-1}$ depending on Galactic 
longitude. 
A rough estimate
of the cosmic background has been obtained by assuming that the
present-day Milky Way is representative of the whole galaxy population
at all times.  With this simple approach, Brocato et al. \cite{BAL98} found
a background number density of $n\sim 10^{-9}$~cm$^{-3}$ and a flux
$c\cdot n\sim$ 40~cm$^{-2}$ s$^{-1}$ (with $c$ the speed of light in vacuum).  
As a matter of fact,
the solar neutrino flux at Earth ($\sim
10^{11}$~cm$^{-2}$ s$^{-1}$) is orders of magnitude larger than
both the Galactic and the cosmic neutrino backgrounds.  

From the experimental point of view, it is presently impossible to
discriminate the non-solar contribution from backgrounds in the detectors.
For this reason, the cosmic and Galactic backgrounds
will remain undetectable until technological advancements will make
possible the construction of highly-efficient directional neutrino
detectors at energies ranging between 0.1 and 15 MeV.

In this paper, the cosmic neutrino background
is obtained by combining the most recent estimates of the 
cosmic star formation history and the stellar initial mass function
with accurate theoretical predictions of the neutrino yields all over the
thermonuclear lifetime of stars of different masses.
We show that, even though this background is hardly of any cosmological 
relevance, its estimated amplitude, 
combined with observations of the $\gamma$-ray 
background, sets an upper limit to the radiative lifetime of neutrinos.
By adopting state-of-the-art models of the structure and evolution of
the Galaxy, 
we also derive maps of the expected flux generated by Galactic sources
as a function of sky direction.
Beyond contributing to the ``noise" level for the detection of single
neutrino sources, these backgrounds are themselves interesting 
since they carry information about the past
history and structure of the Galaxy and our Hubble volume.  Even
though their detection is still beyond present-day technical
capabilities, detailed predictions of their properties
may motivate and inspire future experimental efforts.

The plan of the paper is as follows.
In Section 2 we discuss the cosmic stellar neutrino background.
After introducing the notation and summarizing the basic
properties of thermonuclear neutrino sources, we present our results
and discuss their cosmological relevance and the related uncertainties. 
In Section 3 we estimate the amplitude of the 
Galactic neutrino background and derive maps
of its variation on the sky.
Our conclusions are listed in Section 4.  

\section{Cosmic neutrino background}

\subsection{Method}
Let us consider a homogeneous distribution of isotropic sources of neutrinos
which fills the universe.
Given their comoving neutrino emission rate $\epsilon_\nu(E,z)$ 
(particles per unit time, energy and comoving volume),
the (mean) differential number flux of neutrinos
(particles per unit time, surface,
energy and solid angle) seen by an observer at redshift $z_0$ is 
\be
\bar{I}_\nu^{(z_0)}(E)\equiv
\frac{dN_\nu}{dt\,dS\,dE\,d\Omega}
=\frac{c}{4\pi}
\int_{z_0}^\infty \frac {1+z}{1+z_0}
\,\,\epsilon_\nu\left[E \frac{1+z}{1+z_0},z \right] \left|\frac{dt}{dz}
\right|\, dz\;,
\end{equation}
where 
\be
\left|\frac{dz}{dt}\right|=H_0\, (1+z) 
\,[\Omega_0\,(1+z)^3+\Omega_k\,(1+z)^2+\Omega_\Lambda]^{1/2}\;,
\end{equation}
with $\Omega_k=1-\Omega_0-\Omega_\Lambda$, 
fixes the relationship between cosmic time and redshift.
In agreement with recent estimates (e.g. \cite{TAL03}),
throughout this work we will assume that the present-day
value of the matter density parameter
$\Omega_0=0.3$, the vacuum energy density parameter 
$\Omega_\Lambda=0.7$, and the Hubble 
parameter $H_0=100 \,h\,{\rm km\, s}^{-1}\,{\rm Mpc}^{-1}$
with $h=0.65$. 
It is convenient to separate the energy and redshift dependence in 
$\epsilon_\nu$ by writing,
\begin{equation}
\epsilon_\nu(E,z)=\sum_i S^{(i)}(E)\, R_\nu^{(i)}(z)
\end{equation}
where the index $i$ runs over the different thermonuclear 
channels for neutrino production. In this case, 
$S^{(i)}(E)$ is the spectral energy
distribution of neutrinos at emission (normalized
so that $\int S^{(i)}(E)\,dE=1$), while $R_\nu^{(i)}(z)$ gives the 
rate of neutrinos produced through the $i-$th channel.
This can be estimated by combining a set of basic quantities.
First,
one needs to know when stars have been formed throughout
the universe. This is parameterized by the cosmic star formation rate,
$\psi(z)$, which gives the mass of stars formed per unit time and comoving 
volume as a function of redshift. 
Second, since stars with different masses produce different neutrino rates,
it is important to know their relative abundances.
This is described by
the initial mass function (IMF), $\phi(M)$, which gives the mass 
distribution of stars at birth.
Finally, one needs to know the neutrino emission rate $L^{(i)}_\nu(M,t,Z)$
(particles per unit time) as a function of stellar mass, age, and
chemical composition.
In what follows, we will assume that the neutrino luminosity 
does not depend on the stellar metallicity, $Z$. 
This assumption is supported
by numerical models showing that, independently of the stellar mass, 
$L^{(i)}_\nu$  
varies by less than 10 per cent when the metallicity
is changed from the solar value ($Z_\odot\simeq 0.02$)
to $Z=10^{-4}$ (see Ref. \cite{BAL98} and references therein).
In summary,
the intensity of the background
observed at Earth due to the integrated
effect of all the stellar sources in the universe is  

\begin{equation}
\bar{I}_\nu^{(0)}(E)=
\frac{1}{4\pi}
\frac{c}{H_0} \sum_i \int_0^\infty\!\!\!\! dz\,
\frac{S^{(i)}_\nu[(1+z)E]\,R_\nu^{(i)}(z)}
{[\Omega_0\,(1+z)^3+\Omega_k\,(1+z)^2+\Omega_\Lambda]^{1/2}}\;,
\label{flux}
\end{equation}
where the comoving emission rate of thermonuclear neutrinos is given by
\begin{equation}
R_\nu^{(i)}(z)=
\frac{\int_{M_{\rm
min}}^{M_{\rm max}}\! dM\,\phi(M)\,
\int_{t_M}^{t(z)} dt'\,\Psi[z(t')]\,
L^{(i)}_\nu(M,t-t')}
{\int_{M_{\rm min}}^{M_{\rm max}}\! dM\,M\,\phi(M)}\;,
\label{rate}
\end{equation}
$t(z)$ is the age of the universe at redshift $z$,
and $t_M=\max[0,t(z)-t_{\rm life}(M)]$ with
$t_{\rm life}(M)$ the lifetime of a star of mass $M$ (intended as the time
over which a star is producing energy through H-burning nuclear reactions).

\subsection{Stars as neutrino sources}

The net balance of hydrogen burning is the transformation of 4 protons
and 2 electrons 
into a $^{\mathrm 4}$He nucleus. Conservation of lepton number
requires the emission of 2 electron
neutrinos. The energy
yield of $\sim 27$ 
MeV is mostly released in the form of thermalized photons
in the eV range (corresponding to surface temperatures of $10^3-10^4$ K).
In other words, one neutrino leaves a star every $10^{7}$ photons.
The conversion of hydrogen into helium takes place through different
channels which characterize the neutrino production rates and spectra
for stars
of different mass and age. 
Very-low mass stars ($M \lesssim 0.8 M_{\sun}$) spend more than a
Hubble time in their central hydrogen burning phase.
They convert H into He mainly through the 
proton-proton I (ppI)
chain. Neutrinos are emitted during the fusion of two
protons into a nucleus of deuterium, $p+p \to d+e^++\nu_e$,
(pp neutrinos).
Obviously, these stars had only the chance to produce pp neutrinos during
their past lifetime.
With increasing 
stellar mass, the ppII and ppIII chains
become more efficient.
In consequence, 
low-mass main-sequence (MS) stars emit a mixture of neutrinos generated by
the pp reaction, the electron capture on $^{\mathrm 7}$Be, and
the $^{\mathrm 8}$B  decay. 
Note, however, that, 
in the subsequent H-shell burning phase, the
energy production of these stars is dominated by the CNO cycle.
In this case, also 
neutrinos coming from the $\beta^+$ decay of
$^{\mathrm 13}$N, $^{\mathrm 15}$O, and $^{\mathrm 17}$F 
contribute to the total flux.
In more massive MS stars, the CNO cycle becomes progressively the
dominant mechanism for energy generation, and 
for $M \gtrsim 1.5 M_{\sun}$ CNO neutrinos dominate 
the MS stage.

In this work, we consider all the nuclear reactions which produce
electron
neutrinos in the pp chains and the CNO cycle.
This is done by using the stellar neutrino yields computed by 
Brocato et al. \cite{BAL98}
as a function of stellar mass and age. 
These cover the whole
H and He burning phases for stars with solar chemical composition and
for nine  selected values of the stellar mass ranging from 0.8 to
20~$M_{\sun}$.  
This set of evolutionary tracks covers the three stellar mass ranges
characterized by different evolutionary behaviors related to the
occurrence of electron degeneracy in He stellar cores (low-mass stars)
or in C,O stellar cores (intermediate-mass stars), or to the quiet
ignition of nuclear reactions in 
non-degenerate cores (massive stars).
Regrettably, no numerical model is available for 
$M<0.8  \,M_{\sun}$. However, we found that
for $M<1.7 \,M_{\sun}$,
the neutrino luminosity of pp neutrinos as a function of stellar
mass is very well described by a power-law scaling relation $\propto M^{4.4}$.
We use this function to extrapolate
the emission rate of pp neutrinos down to 0.08 $M_{\sun}$
(the minimum mass for hydrogen burning stars). 

While neutrino emission rates depend on the star structure,
neutrino spectra are derived from the standard theory of
electroweak interactions and are largely independent of stellar 
parameters \cite{BAH91}\footnote{In principle, neutrino energy spectra
are only slightly dependent on the temperature at which the nuclear reactions 
take place \cite{BAH94}.}.
For pp and CNO neutrinos we use the energy distributions obtained in
Refs. \cite{BAH97} and \cite{BU88}, respectively.
The line shapes for the neutrinos emitted by the
electron-capture process which causes the transition between the 
ground-state of 
$^{\mathrm 7}$Be to either the ground-state or the excited state of 
$^{\mathrm 7}$Li are taken from Ref. \cite{BAH94}.
For the high-energy tail of the neutrino spectrum due to
$^{\mathrm 8}$B decay we use the results presentend in Ref. \cite{BAH96}.

In what follows, we will only consider electron neutrinos produced 
by thermonuclear reactions burning H and He. 
In the advanced phases of
stellar evolution, additional neutrinos and anti-neutrinos, of all three
flavors, can be emitted directly at expenses of the thermal energy of
the star (the so-called ``cooling'' neutrinos). They have energies
between 10 and 60 KeV \cite{SAL87},
and their expected emission rates are comparable
to those of thermonuclear neutrinos \cite{BAL98}. For this
reason we will limit our discussion to energies larger than 0.06 MeV.
Oxygen and silicon burnings in massive stars
(which respectively proceed at a temperature of about 160 and 270 keV) 
can easily produce thermal neutrinos up to the MeV scale. 
These processes are active for a short time
(their duration is of the order of 1 yr for stars of 25 $M_\odot$)
and liberate a lot less binding energy per nucleon than hydrogen
burning, so that the total energy release from such phases is
significantly smaller. 
However, at variance with hydrogen burning,
almost all of the energy released from these phases goes into
neutrinos. 
Also, Type Ia supernovae are powerful neutrino sources in the
sub-MeV range while, above a few MeV, core-collapse neutrinos are
expected to dominate the counts. 
We defer the analysis of the background contribution of these sources to 
future work.

\begin{figure}[b]
\centerline{\includegraphics[height=8.0cm]{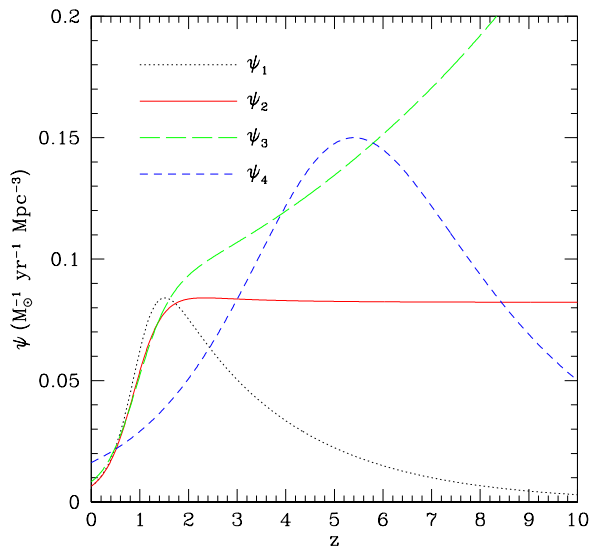}} 
\caption{Cosmic star formation history as a function of redshift.
Different linestyles correspond to the functions listed in equation
(\ref{SFR}).}
\label{SFR_fig}
\end{figure}

\subsection{Star formation history}

In the past few years a number of authors have attempted to determine
the star formation history of the universe from observational data.
Although some details remain controversial,
the evolution of the star formation rate (hereafter SFR) with redshift
can be traced till to $z\sim 6$ using various diagnostics.
It is widely accepted that for $z<1$ the SFR is a rapidly increasing function
of redshift, reflecting the fact that present-day galaxies have been
assembled long ago.
Corrections for dust reddening and extinction complicate the picture at 
higher redshifts.
To summarize the present knowledge and the corresponding uncertainty,
we use here four different parameterizations of the global star-formation
rate per unit comoving volume:
\begin{eqnarray}
\label{SFR}
\psi_1(z)&=&0.3 \, \frac{h}{0.65}\, \frac{\exp(3.4\,z)}{\exp(3.8\,z)+45}
\,\Upsilon(z)
\,M_\odot\,{\rm yr}^{-1}\,{\rm Mpc}^{-3}\;,\nonumber\\
\psi_2(z)&=& 0.15 \,\frac{h}{0.65} \,\frac{\exp(3.4\,z)}{\exp(3.4\,z)+22}
\,\Upsilon(z)
\,M_\odot\,{\rm yr}^{-1}\,{\rm Mpc}^{-3}\;,\\
\psi_3(z)&=& 0.2 \,\frac{h}{0.65} \,\frac{\exp(3.05\,z-0.4)}{\exp(2.93\,z)+15}
\,\Upsilon(z)
\,M_\odot\,{\rm yr}^{-1}\,{\rm Mpc}^{-3}\;,\nonumber\\
\psi_4(z)&=& 0.15\, \frac{14 \exp[0.6 (z-5.4)]}{5+9\exp[0.933(z-5.4)]}
\,M_\odot\,{\rm yr}^{-1}\,{\rm Mpc}^{-3}\;.\nonumber
\end{eqnarray}
with
\be
\Upsilon(z)=\frac{[\Omega_0\,(1+z)^3+\Omega_k\,(1+z)^2+\Lambda]^{1/2}}{(1+z)^{3/2}}\;.
\end{equation}
The first three, obtained by Porciani \& Madau \cite{PM01}, 
match present data
from UV-continuum and H-$\alpha$ cosmic luminosity densities,
and allow for different
amounts of dust reddening, especially at high-$z$ (see Ref. \cite{PM01}
for further details). The fourth SFR is taken from 
Ref. \cite{SH03} 
(see also \cite{NAL01}) 
and is based on theoretical speculation only. This is obtained from a 
set of cosmological simulations including gas hydrodynamics where an a priori 
recipe for star formation has been assumed. The whole analysis has been
developed in the framework of a vacuum energy dominated cold dark matter model
in which galaxies form hierarchically. 
At variance with the determinations based on observational data,
this SFR  extends to very high redshifts but suffers from a number of
theoretical assumptions. The resulting function has a maximum at $z\sim 5.5$
and declines roughly exponentially towards both high and low redshifts.
A semianalytic model to explain the origin of this functional form
has been presented by Hernquist \& Springel \cite{HS03}.
The SFRs given in equation (\ref{SFR}) are plotted in Figure \ref{SFR_fig}. 
The intensity of the cosmic neutrino background have been
computed by integrating equation (\ref{flux}) out to $z_{\rm max}=10$.

\subsection{Initial Mass Function}
Basic reasoning suggests that the IMF should vary with the properties
of the star-forming clouds. Nevertheless, although there is no fundamental
reason for a universal mass distribution of stars, 
so far no firm evidence for a variable IMF exists
(see e.g. Refs. \cite{EIS01,GIL01,KRO02}
for different
points of view regarding this issue).
In what follows we will assume a time independent IMF, adopting different
functional forms to take into account observational uncertainties.
As a standard model we use the IMF originally proposed
by Salpeter: $\phi(M)\propto M^{-2.35}$.
Even though this is supported by many studies,
especially for $M\gtrsim 0.5\,M_{\odot}$ 
(see e.g. \cite{MJD95,KRO01}),  
the behavior of the IMF at lower masses is still very
uncertain 
(see e.g. \cite{KRO01,SCA98,RR02}). 
This reflects the observational uncertainties
in the determination of the star luminosity function at low 
luminosities. 
To represent a number of different observational results, we focus here
on two additional examples.
From M-dwarf observations with the Hubble Space
Telescope, Gould, Bahcall \& Flynn \cite{GBF96} determined the following
form for the low-mass end of the IMF ($M<1.6 M_{\odot}$):
\begin{equation}
\log (\phi) =  {\rm const.}-2.33 \log (M/M_{\odot}) -1.82 \ [\log
(M/M_{\odot})]^{2}\;.
\label{gould}
\end{equation}
This function peaks at $M\sim 0.23\, M_{\odot}$, and the slope at $M =
1\, M_{\odot}$ is $-2.33$, almost coincident with the Salpeter value of
$-2.35$.
We will call GBF IMF the distribution obtained by matching
this equation for $M<1 \,M_{\odot}$ with a
Salpeter IMF for higher masses.
Finally, as a third option, we use a parameterization of the IMF 
obtained from star counts in different Galactic fields by
Kroupa \cite{KRO01}.
This is well fit by a broken power-law with a change in the exponent near 
$0.5 \,M/M_{\odot}$, 
namely, $\phi \propto M^{-\alpha_{i}}$, with
$ \alpha_{1} = 1.3$ for  $M/M_{\odot} \le 0.5$ and 
$\alpha_{2}  = 2.3$ for $M/M_{\odot} > 0.5$.
In all cases, we assume $M_{\rm min}=0.08 \,M_\sun$ and $M_{\rm max}=125
\,M_\sun$.

\subsection{Results}
\begin{figure}[thb]
\centerline{\includegraphics[height=8.0cm]{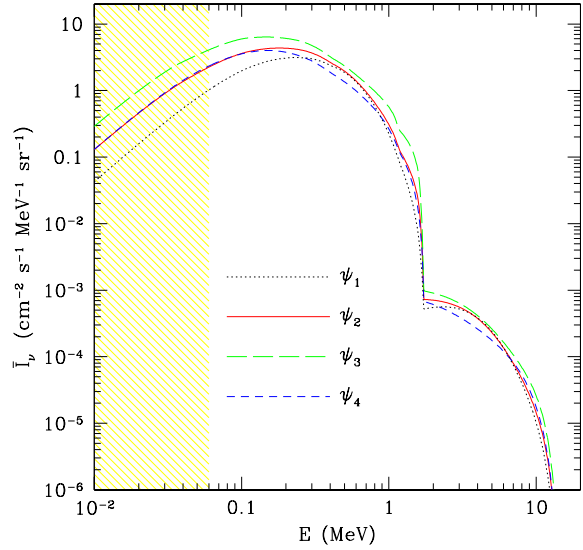}} 
\caption{Mean intensity of the cosmic neutrino background as a function of 
energy.
Different linestyles correspond to different assumed SFRs.
We only consider neutrinos produced by thermonuclear reactions in stars. 
The shaded region highlights the energy range in which additional
``cooling neutrinos'' are expected to contribute significantly.}
\label{nu_sfr}
\end{figure}

In this section, we present
our main results, obtained by combining in equation (\ref{flux})
the different functions discussed above.
In Figure \ref{nu_sfr}, 
the expected cosmic background intensity, $\bar{I}_\nu^{(0)}$, 
is plotted as a function of energy
for different star formation histories and assuming a Salpeter IMF.
The corresponding fluxes, integrated over all energies, are listed in 
Table \ref{tbl-1}.\footnote{
In the literature,
it has become customary to call flux the quantity
$c\, n\equiv 4\pi \bar{I}_\nu$ even though this does not correspond to any 
physical flux. In fact, for isotropically moving particles the net flux vector
$\langle n \,{\bf v}\rangle$ vanishes, while the (one-sided)
flux through a fixed surface is given by
$
F_\nu^{+}=\int_{2 \pi} I_\nu(\theta,\phi)\,\cos\theta\,d\Omega,
$
which reduces to $F_\nu^{+}=\pi\,I_\nu$ in the isotropic case.} 
\begin{figure}[b]
\centerline{\includegraphics[height=8.0cm]{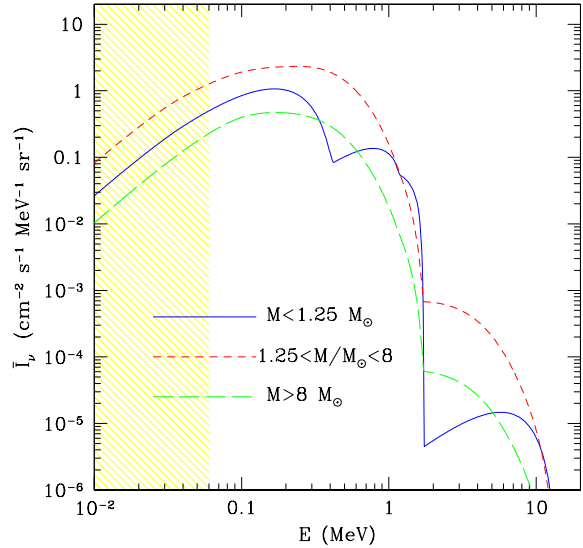}} 
\caption{Contributions to the
neutrino differential flux due to stars of different masses.
A Salpeter IMF and $\psi_2(z)$ are assumed.}
\label{nu_mass}
\end{figure}
The amplitude of the background ranges within a factor of 2, with $\psi_3$ and
$\psi_1$ giving the highest and the lowest fluxes, respectively.
This is not surprising since $\psi_1$ has a much lower amplitude than the
other SFRs at $z>2$, while $\psi_3$ keeps increasing with redshift. 
It is interesting to compare the contributions to the background due
to low ($M/M_\odot<1.2$), intermediate ($1.2\le M/M_\odot < 8\,M_\odot$) 
and high mass stars ($M/M_\odot\ge 8$).
This is done in  Figure \ref{nu_mass}, where  a Salpeter IMF and 
$\psi_2(z)$ are assumed. 
The shape of the different curves can be explained as follows.
Very-low mass stars, $M/M_{\sun}\lesssim 0.9$, produce only pp neutrinos
with energies below 0.4 MeV.
\begin{deluxetable}{ccccccccc}
\tabletypesize{\scriptsize}
\tablecaption{Cosmic neutrino background. \label{tbl-1}}
\tablewidth{0pt}
\tablehead{
\colhead{SFR} & \colhead{IMF}   & \colhead{$c\,n$}
&\colhead{$n\cdot10^9$}&
\colhead{$u\cdot10^{10}$}  & 
\colhead{$\Omega_\nu\,h^2\cdot 10^8$} & 
\colhead{$\bar{E}$} &
\colhead{$\bar{z}$\tablenotemark{a}}&
\colhead{$s\cdot 10^7/k_{\rm b}$}
\\
& &  \colhead{$({\rm cm}^{-2}\,{\rm s}^{-1})$}&
\colhead{$({\rm cm}^{-3})$}&
\colhead{$({\rm MeV}\,{\rm cm}^{-3})$}&
\colhead{} 
&\colhead{(MeV)}&
&\colhead{$({\rm cm}^{-3})$}}
\startdata
1 & Salp. & 20.2 & 0.68 & 2.73 & 2.59 & 0.40 & 1.41 &0.61 \\
1 & Kroupa& 29.8 & 0.99 & 4.03 & 3.82 & 0.40 & 1.45 &0.89 \\
1 & GBF   & 35.9 & 1.20 & 4.86 & 4.61 & 0.41 & 1.43 &1.07 \\
\\
2 & Salp. & 26.0 & 0.87 & 3.29 & 3.12 & 0.38 & 1.83 &0.77  \\
2 & Kroupa& 38.2 & 1.27 & 4.81 & 4.57 & 0.38 & 1.88 &1.13  \\
2 & GBF   & 46.1 & 1.54 & 5.85 & 5.55 & 0.38 & 1.86 &1.36  \\
\\
3 & Salp. & 36.7 & 1.22 & 4.67 & 4.43 & 0.38 & 2.01 &1.09  \\
3 & Kroupa& 53.5 & 1.79 & 6.78 & 6.43 & 0.38 & 2.06 &1.58  \\
3 & GBF   & 65.0 & 2.17 & 8.32 & 7.90 & 0.38 & 2.04 &1.92  \\
\\
4 & Salp. & 21.2 & 0.71 & 2.56 & 2.43 & 0.36 & 2.30 &0.63  \\
4 & Kroupa& 31.3 & 1.04 & 3.78 & 3.59 & 0.36 & 2.36 &0.93  \\
4 & GBF   & 37.6 & 1.26 & 4.57 & 4.34 & 0.36 & 2.33 &1.11  \\
\enddata
\tablenotetext{a}{This denotes the mean emission redshift for
background neutrinos with $E=0.2$ MeV.}
\end{deluxetable}
Masses in the range $0.9 \lesssim M/M_{\sun} \lesssim 1.2$ pass
through the ppII and ppIII chains, thus emitting $^{\mathrm 7}$Be and
$^{\mathrm 8}$B neutrinos. The small contribution of $^{\mathrm 8}$B
neutrinos to the total flux is evident at higher energies. 
The peculiar feature at energies in the range $0.4 \lesssim (E/ 
1\ \mathrm {MeV}) \lesssim 2$ is due to the reduced quantity of CNO
neutrinos emitted by these stars during their H-burning shell phase.
As discussed above, the stellar neutrino luminosity rapidly increases
with mass and, in particular, the contribution of CNO, $^{\mathrm
8}$B, and $^{\mathrm 7}$Be neutrinos increases. Nevertheless, the steep
slope of the IMF and the shorter main sequence lifetime diminish the 
contribution of massive stars to the stellar neutrino background.
\begin{figure}[t]
\centerline{\includegraphics[height=8.0cm]{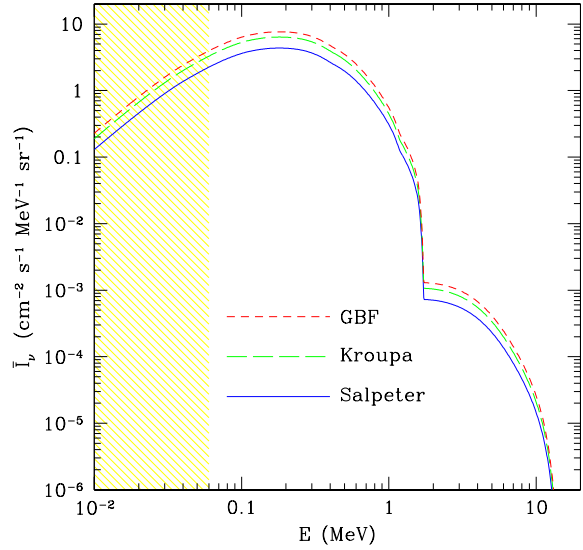}} 
\caption{As in Fig. 1 but for different IMFs, assuming $\psi_2$ for the SFR.}
\label{nu_imf}
\end{figure}
In summary, 
intermediate mass stars clearly dominate the background at all energies
but for a narrow window around 1.5 MeV where small mass stars give a
slightly larger contribution.
In Figure \ref{nu_imf} we show how the spectrum of the background changes
by varying the IMF while using the same SFR (in this case, $\psi_2$).
The reader is referred to Table \ref{tbl-1} for the corresponding
fluxes.
The net effect is a slight spectral distortion and a difference in the 
background intensity as large as a factor of 2. As expected, the amplitude
of the background increases when the relative weight of low-mass stars
is reduced.
\begin{figure}[b]
\centerline{\includegraphics[height=7.5cm]{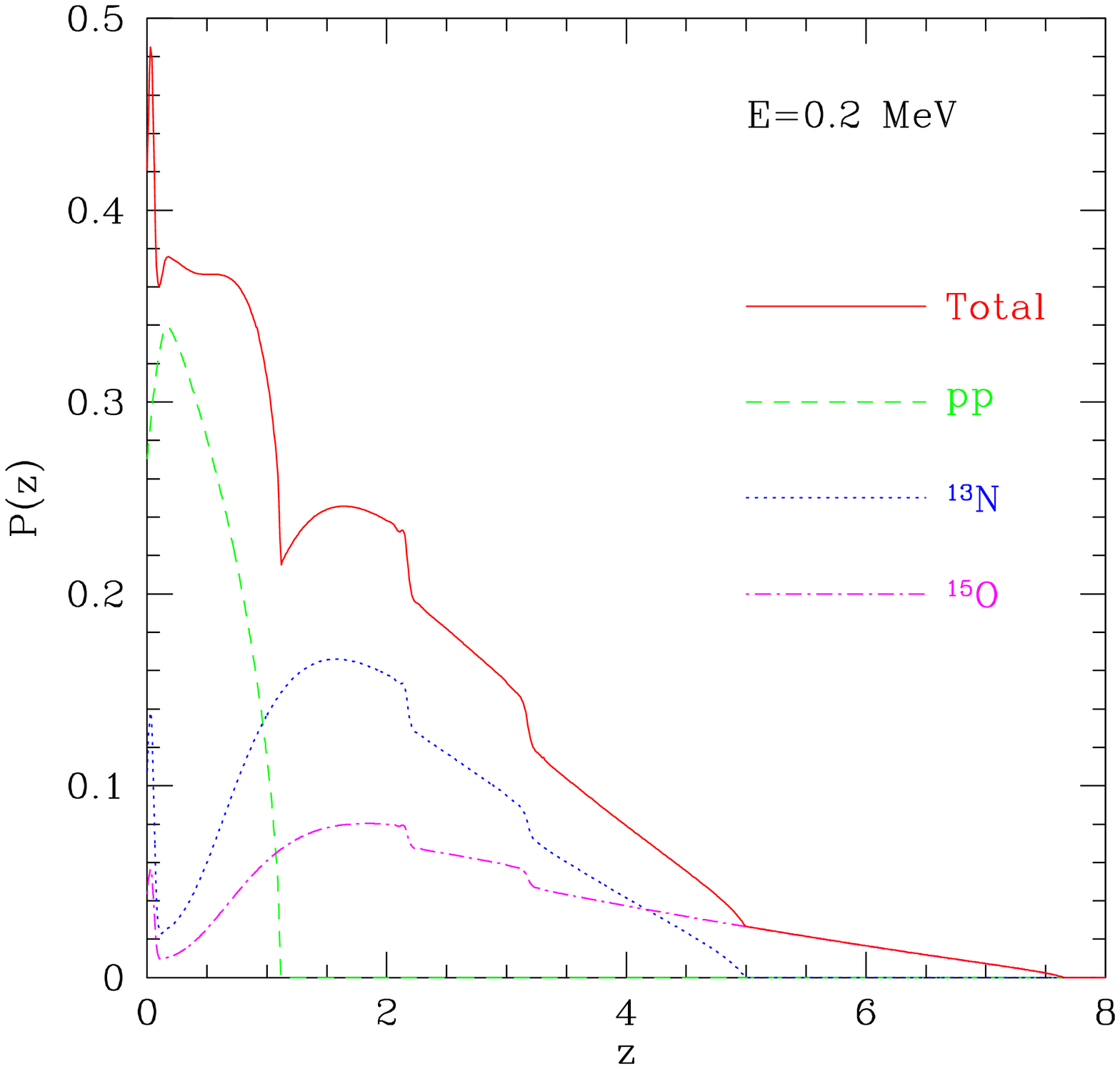}
\includegraphics[height=7.5cm]{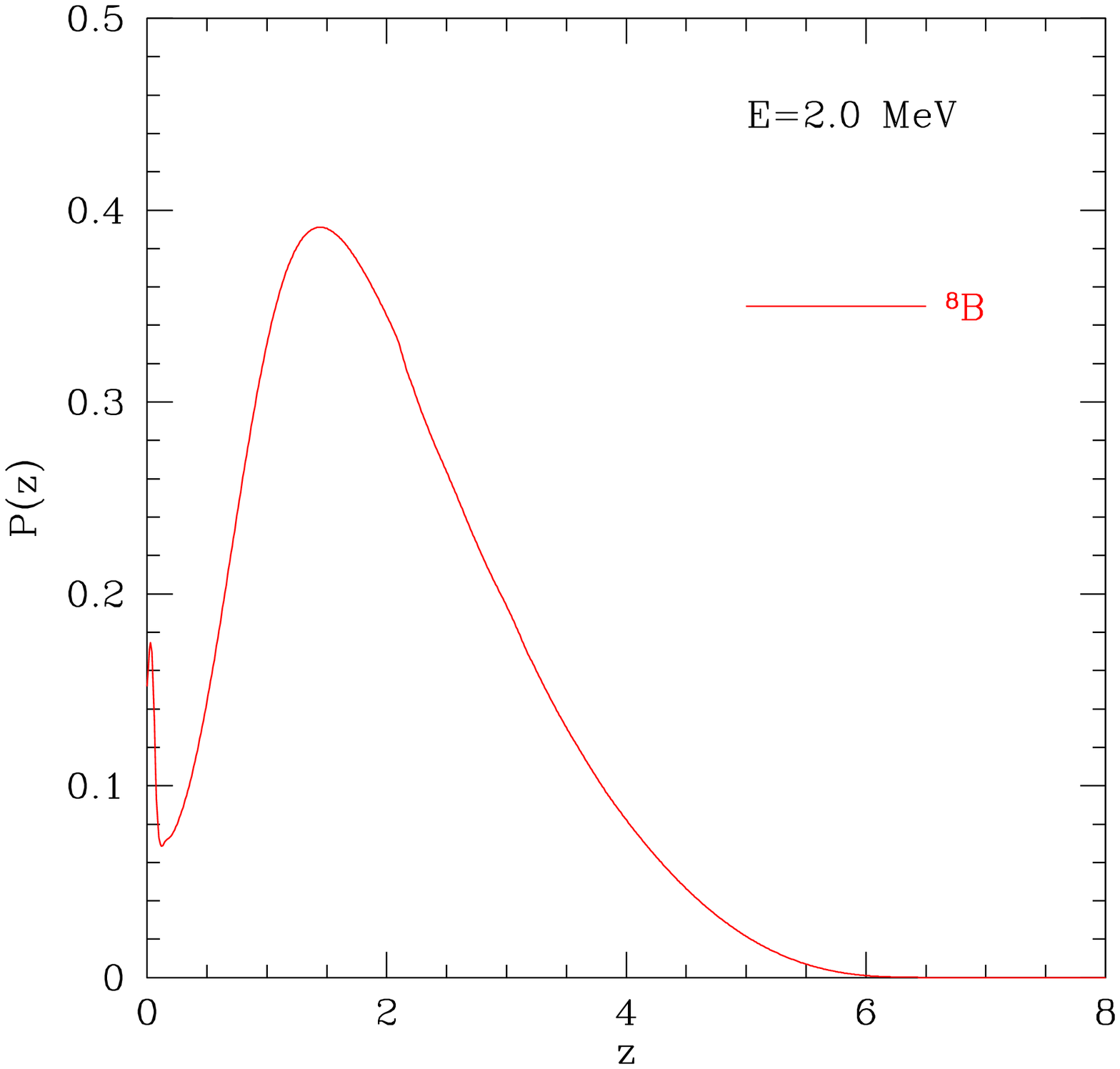}} 
\caption{Probability distribution of the emission redshift for neutrinos
with energy $E$ in the cosmic background (solid line).
Different linestyles denote the contribution of different generation
processes, namely: pp fusion (dashed), $^{\mathrm 13}$N decay
(dotted), and $^{\mathrm 15}$O decay (dot-dashed).
Left and right panels refer to $E=0.2$ MeV and $E=2$ MeV, respectively.}
\label{imf_z}
\end{figure}

It is interesting to study how the 
the probability density distribution for the emission redshift of neutrinos
varies with their energy at detection. 
This is simply given by
\be
P(z,E)=
\frac{c}{4\pi}
\frac{
\epsilon_\nu\left[E (1+z),z \right]}{H_0\,[\Omega_0(1+z)^3+\Omega_k
(1+z)^2+\Omega_\Lambda]^{1/2}} \frac{1}{I_\nu^{(0)}(E)}
\;,
\end{equation}
and it is shown in Figure \ref{imf_z} for two different neutrino energies.
Note that neutrinos at the peak of the energy spectrum ($E\simeq 0.2$ MeV)
have been emitted within a wide redshift range. 
Three different generation mechanisms contribute: the nuclear fusion of 2 
protons 
(which accounts for the 27 per cent of the neutrinos observed at $E=0.2$ MeV)
and the $\beta^+$ decays
of $^{\mathrm 13}$N (44 per cent) and $^{\mathrm 15}$O (29 per cent). 
Each of these processes corresponds
to a different distribution of emission redshifts, which is determined
by the joint effect of the SFR and the shape of the rest-frame emission 
spectrum. For instance,  
the narrow peak near $z=0$ is generated by stars with $M\simeq 1 M_\odot$
which started to produce energy through the CNO cycle only recently.
We find that the mean emission redshift for photons at $E=0.2$ MeV is 
$\bar{z}=1.83$. 
On the other hand, at higher energies, only the $^{\mathrm 8}$B $\beta^+$ 
decay contributes. The emission-redshift distribution for neutrinos
at $E=2$ MeV is shown in the right panel of Figure \ref{imf_z}. 
In this case, we find $\bar{z}=2.03$.

For $E<0.06$ MeV, the contribution to the total
neutrino fluxes of cooling neutrinos should eventually be taken into
account. In fact, the estimated neutrino fluxes at the Earth's surface
due to plasma-neutrinos emitted by stars till the end of the Red Giant
phase as well as during the last phase of cooling as white dwarfs is
comparable to the pp-neutrinos flux \cite{BAL98}.

\subsection{Cosmological relevance}
It is interesting to compare  the energy and entropy densities of  
neutrinos generated
by thermonuclear reactions in the universe with the corresponding
quantities for the primordial neutrino background.
The present-day number and energy densities of the cosmic neutrino background 
generated by stars are 
\be
n=\frac{4\pi}{c}\int \bar{I}_\nu^{(0)}(E)\, dE\ \ \ \ \ \ {\rm and}\ \ \
\ \ \ 
u=\frac{4\pi}{c}\int E\, \bar{I}_\nu^{(0)}(E)\, dE\;,
\end{equation}
respectively.
We define the entropy density for our non-thermal cosmic neutrino background as
\be
s=\frac{4\pi\,k_{\rm b}\,g_{\rm s}}{c^3\,h^3} \int_0^\infty E^2\left\{
f(E) \ln\left[\frac{1}{f(E)}-1\right]
-\ln\left[1-f(E)\right] 
\right\} dE\;,
\label{entropy}
\end{equation}
where 
\be
f(E)=h^3\,c^2\, \frac{\bar{I}_\nu^{(0)}(E)}{E^2}
\simeq 6.36\times 10^{-41} ({\rm MeV}^3\,{\rm cm}^2\,{\rm s})
\,\frac{\bar{I}_\nu^{(0)}(E)}{E^2} \;,
\end{equation}
$k_{\rm b}$ denotes the Boltzmann constant,
$g_{\rm s}$ is the number of helicity degrees of freedom,
and $h$ is the Planck's constant.
This is derived from first principles in the Appendix and gives the standard
result $s= 4.202 \,k_{\rm b}\, n$ for a thermal background.
Note that, since neutrinos in the cosmic background are far from being 
degenerate (i.e. $f(E)\ll 1$), equation
(\ref{entropy}) reduces, in practice, to 
$s\propto -\int E^2 \,f(E)\, \ln[f(E)] \,dE$. 
Values for the number, energy and entropy densities of the background are 
listed in Table \ref{tbl-1} for the different models we considered. 
Typically, $n\simeq 10^{-9}\,{\rm cm}^{-3}$, $u\simeq 5\times 10^{-10}
\,{\rm MeV\, cm}^{-3}$ (corresponding to a contribution to the
density parameter of $\Omega_\nu\,h^2\simeq 
5\times 10^{-8}$)
and $s\simeq 10^{-7} \,k_{\rm b}\,{\rm cm}^{-3}$.
On the other hand,
from the standard theory of neutrino decoupling in the early universe one
expects a thermal distribution of relic particles with $T_{\rm rel}=1.95$ K.
This contributes
$n_{\rm rel}=56.5 \,{\rm cm}^{-3}$, 
$\Omega_{\rm rel}\,h^2\simeq 4.8\times 10^{-6}$ (for massless neutrinos), 
\footnote{Assuming that the 
sum of neutrino masses amounts to, at least, 1/30 eV
(as suggested by recent data) 
corresponds to $\Omega_{\rm rel}\,h^2\gtrsim 3.5\times 10^{-4}$.}
and $s_{\rm rel}=203.5 \,k_{\rm b}\,{\rm cm}^{-3}$ for each neutrino family
(an equal contribution is given by the corresponding antiparticles).
We can thus draw the conclusion that thermonuclearly produced neutrinos
hardly are of any cosmological relevance. In particular, their abundance and
entropy are overwhelmed by the corresponding properties of relic
neutrinos from the hot initial phases of the universe. 
However, since thermonuclear neutrinos from stars are characterized by much
higher energies than the thermal relics from the big bang,
the ratio of the energy densities of this two 
backgrounds is much smaller than the number density ratio. 
In both cases, the contribution to the 
curvature of space-time is anyway negligible.

\subsection{Constraints on the neutrino radiative lifetime
and electromagnetic form factors}

In the hypothesis of radiative neutrino decay, $\nu \rightarrow 
\nu_{\ell} + \gamma$, cosmic neutrinos ($\nu$ with mass $m$) 
decaying into some neutral fermion (possibly a lighter neutrino
$\nu_{\ell}$ with mass $m_\ell$) over a Hubble time
would be the source of a diffuse photon background. 
For a fixed energy 
$E_{\nu}$ of the heavy neutrino, the photon spectrum is continuum, with 
an end point at $E_{\gamma}= E_{\nu} (m^2-m_{\ell}^2)/
m^2$ for ultrarelativistic neutrinos. 
Depending on the mass pattern, the photon 
energy can extend up to the MeV region. 
As derived in Appendix B, the background intensity of the 
resulting photons at $z=0$ is
\be
j(E)=\frac{m \,c^2}{\tau_\gamma}\,
\frac{\alpha(E)}{H_0\,E}\,\bar{I}_\nu^{(0)}(2E)\;,
\end{equation}
where  $\tau^{-1}_{\gamma}$ is the radiative decay rate and 
\be
\alpha(E)=
\int_0^\infty \!\!\! \frac{dz}{(1+z)^2\,
[\Omega_0(1+z)^3+\Omega_k(1+z)^2+\Omega_\Lambda]^{1/2}}
\int_z^\infty \!\!\! dz'\,P(z',2E)
\;,
\end{equation}
a numerical coefficient of order unity.

Observations have shown the presence of a diffuse (high-latitude)
photon background in the MeV region 
that appears to be of extragalactic origin.
Both a large number of unresolved point sources and truly diffuse
processes might potentially contribute. However,
the X-ray background (below 100 keV) is generally understood to
arise primarily from the integrated emission of active galactic 
nuclei (such as Seyfert galaxies) with redshifts ranging up to 6
\cite{LB82}. 
At slightly higher energies ($E\sim 1$ MeV), 
radioactivity in Type Ia supernovae is expected to be the most important
contributor \cite{WAL99}.

The amplitude of the diffuse $\gamma$-ray background in the MeV 
range is extremely hard to measure (both from the Earth and from space)
due to the extraordinarily high atmospheric and instrumental background. 
In particular, the interaction of cosmic rays with the detector 
and nuclear line emission dominate the noise. 
Recent results from a number of satellite missions indicate the presence of
an extragalactic component significantly lower than previous measurements.
The Medium Energy Detectors of the A4 experiment onboard the 
High Energy Astrophysics Observatory 1 (HEAO-1) measured a
background intensity
$j(E)=(2.62\pm 0.05)\times(E/0.1\,{\rm MeV})^{-2.75\pm 0.08}\,
{\rm cm}^{-2}\,{\rm s}^{-1}\,{\rm MeV}^{-1}\,{\rm sr}^{-1}$ 
for $0.1\leq (E/1\, {\rm MeV}) \leq 0.4$ \cite{KAL96}.
The Imaging Compton Telescope (COMPTEL) on the Compton Gamma Ray Observatory 
(CGRO) showed that the diffuse background at
$0.8\leq (E/1\, {\rm MeV})\leq 30$ 
is consistent with power-law extrapolation from the HEAO result at lower 
energies \cite{KAL96}, and provided a new fitting formula 
$j(E)=(1.05 \pm 0.2)\times 10^{-4}\,(E/5\, {\rm MeV})^{-2.4\pm 0.2}
\,{\rm cm}^{-2}\,{\rm s}^{-1}\,{\rm MeV}^{-1}\,{\rm sr}^{-1}$
\cite{KAP98}. 
Finally,
the Solar Maximum Mission $\gamma$-ray spectrometer (SMM/GRS)
measured a signal for $0.3\leq (E/1\, {\rm MeV})\leq 8.5$
which is $\sim 20\%$ lower than
the COMPTEL determination \cite{WAL96}.

If one attributes a fraction $f$
of the diffuse background to cosmic neutrino radiative decay, one can 
derive a lower bound for the radiative decay lifetime, $\tau_\gamma$
(the total lifetime is obtained by multiplication with the branching
ratio for the radiative channel), 
\footnote{This limit is obtained for photons of 0.1 MeV, corresponding
to neutrinos at the peak of the cosmic spectrum. On the grounds of the
recent experimental results on flavor mixing and oscillation, this
lower bound is valid for any neutrino type.}
\be
\frac{\tau_\gamma}{m\,c^2} \geq \frac{(2-6)}{f}\times 10^{12}\, 
\,{\rm s}\,\,{\rm eV}^{-1}\;.
\end{equation}
The radiative decay rate can be uniquely characterized by the magnetic
and electric (dipole) transition moments, 
$\mu_{\nu\, \nu_\ell}$ and $\epsilon_{\nu\,\nu_\ell}$,
through (e.g. \cite{MP91})
\begin{equation}
\label{dec}
\frac{1}{\tau_\gamma}=\frac{\mu_{\rm eff}^2}{8\pi}
\;\,m^3\,\left(1-\frac{m_{\ell}^2}{m^2}\right)^3\;, 
\end{equation}
where $\mu_{\rm eff}^2=|\mu_{\nu\, \nu_\ell}|^2+|\epsilon_{\nu\, \nu_\ell}|^2$
and natural units ($\hbar=c=1$) have been adopted for convenience.
Assuming a mass hierarchy $m\gg m_\ell$ and measuring
the effective magnetic moment
in units of the Bohr magneton, $\mu_{\rm B}$, this corresponds to 
\begin{equation}
\frac{\mu_{\rm eff}}{\mu_{\rm B}}\leq
(1-4)\times10^{-7}\,f^{1/2}\,\left(\frac{m\,c^2}{\rm 1\ eV}\right)^{-2}
\;.
\end{equation}
Note that, for $m\,c^2\ll 1$ eV (the favored mass range from current
experiments), the radiative decay channel is vastly phase-space suppressed,
and the phase-space factor $m^3$ in equation 
(\ref{dec}) strongly reduces the strength of the limit on $\mu_{\rm eff}$.
Anyway, even by assuming $f=1$,
our upper bound is much more stringent than 
the limit $\mu_{\rm eff}/\mu_{\rm B}\leq
0.09 \,(m\,c^2/1\ {\rm eV})^{-2}$ derived from 
the absence of decay photons of reactor $\bar{\nu}_e$ fluxes
\cite{OvM87}, and 
from the solar neutrino flux, 
$\mu_{\rm eff}/\mu_{\rm B}\leq
5\times10^{-6} \,(m\,c^2/1\ {\rm eV})^{-2}$
\cite{RAF85}.
However, most of the hard X-ray background (2-10 keV)
has been recently resolved into discrete sources \cite{MAL00}.
The data compilation by Moretti et al. \cite{MAL03} shows that 90-95\% of the
background can be ascribed to discrete source emission.
Accounting for scattered light, the resolved fraction might even equal
99\% of the total \cite{SOL03}.
Active galactic nuclei may similarly contribute 
even at 100 keV \cite{ZZK93}.
Assuming $f\simeq0.01$, makes our result comparable to the limit
derived from the Supernova 1987A neutrino burst,
$\mu_{\rm eff}/\mu_{\rm B}\leq
1.5 \times10^{-8} \,(m\,c^2/1\ {\rm eV})^{-2}$,
(see Ref. \cite{OAL93} and references therein).
\footnote{For Dirac neutrinos,
much more restrictive limits on the magnetic moments
can be derived from Supernova 1987A data
\cite{BM88}.}

Our bound on $\mu_{\rm eff}$, however, is not competitive with
other astrophysical determinations (see \cite{RAF99} for 
a comprehensive review)
based, for instance, on plasmon decay in globular cluster stars
($\mu_{\rm eff}/\mu_{\rm B}\leq 3\times 10^{-12}$) 
\cite{HRW94}, the observation of TeV $\gamma$-rays from Markarian 421 and 501
\cite{BIL98}, helioseismology ($\mu_{\rm eff}/\mu_{\rm B}\leq
4\times 10^{-10}$) \cite{RAF99}, 
and the primordial neutrino background ($\mu_{\rm eff}/\mu_{\rm B}\leq
10^{-11} \,(m\,c^2/1\ {\rm eV})^{-9/4}$) \cite{RT90}.
Laboratory limits on the diagonal and transition magnetic moments
are typically obtained from measurements of the $\bar{\nu}\,e$ 
scattering cross section at low energy near a nuclear reactor.
Current upper limits to the transition moments involving a particular
flavor are:
$\mu/\mu_{\rm B}\leq 1.0\times 10^{-10}$ for $\nu_e$ 
\cite{DAL03}, 
$\mu/\mu_{\rm B}\leq 7.4\times 10^{-10}$ for $\nu_\mu$ 
\cite{KAL90},
and $\mu/\mu_{\rm B}\leq 5.4\times 10^{-7}$ for $\nu_\tau$ 
\cite{CAL92}.

\section{Galactic neutrino background}
Will we ever be able to detect the cosmic neutrino background 
generated by stars directly?
Beyond technological issues related to the availability of detectors with
the required sensitivity, one needs to worry about the presence of other
backgrounds. 
It is straightforward to think of neutrinos
generated by thermonuclear reactions taking place within the Galaxy, since
it is expected to have a similar spectrum to the cosmic one.
In this section we compute the basic properties of this new background and 
its directional dependence on the sky. 
For analytical convenience, we will use here a different parameterization 
of the star formation rate, $\widetilde{\psi}^{(j)}(t')$, denoting
the mass of gas converted into stars per unit time in a given
galatic component (thin disk, thick disk, halo and bulge) labelled
by the index $(j)$.
The mean density of stars that are still thermonuclearly active today can
then be expressed as
\be
n_{\rm active}=
\frac{\int_{M_{\rm min}}^{M_{\rm max}}\! dM\,\phi(M)\,
\int_{t_M}^{t_0} dt'\,\widetilde{\psi}^{(j)}(t')}
{\int_{M_{\rm min}}^{M_{\rm max}}\! dM\,M\, \phi(M)}\;,
\end{equation}
where $t_0=14.5$ Gyr is the present cosmic age in the adopted cosmological 
model.
Therefore,
the mean neutrino rate presently emitted by a star in the component $(j)$
through the production channel $(i)$ is 
\begin{equation}
G_\nu^{(i,j)}=
\frac{\int_{M_{\rm min}}^{M_{\rm max}}\! dM\,\phi(M)\,
\int_{t_M}^{t_0} dt'\,\widetilde{\psi}^{(j)}(t')\,
L^{(i)}_\nu(M,t_0-t')}
{\int_{M_{\rm min}}^{M_{\rm max}}\! dM\,\phi(M)
\, \int_{t_M}^{t_0} dt'\,\widetilde{\psi}^{(j)}(t')
}\;,
\label{grate}
\end{equation}
where, once again, $t_M=\max[0,t(z)-t_{\rm life}(M)]$.
Note that, being a mean quantity, $G_\nu^{(i,j)}$ does not depend on the 
overall normalization of the star-formation rate but only on its temporal 
evolution.
We account for the geometric structure of the galaxy by introducing
a set of functions, $n^{(j)}({\bf r}_{\rm g})$, 
which describe the present-day number 
density distributions of stars in the different galactic components
as a function of the distance from the galactic centre. 
The differential neutrino flux detected on Earth is thus 
\begin{equation}
I_\nu^{(0)}(\hat{\bf r},E)=\frac{1}{4 \pi}\sum_{i,j} G_\nu^{(i,j)}\,
S^{(i)}_\nu(E) \,
\int dr\,
n^{(j)}({\bf r}+{\bf r}_{0})\;,
\label{gflux}
\end{equation}
where ${\bf r}$ denotes separations from Earth and ${\bf r}_0$ is
our distance from the galactic centre.

In the last decades,
accurate star count surveys, gas-dynamical and stellar-kinematics studies,
and the analysis of near-infrared brightness maps
allowed us to identify the main structural features of the Milky Way
and to formulate a number of standard density laws to describe its
populations. 
Recently, a series of technological improvements made possible
a more accurate determination of the corresponding free parameters.
In this section, we will use these results to estimate the intensity and 
directional dependence of the Galactic neutrino background.
In particular, for the outer galaxy, we will use the results from
the high-latitude star counts obtained by the Sloan Digital Sky Survey
\cite{CAL01} and by Siegel et al. \cite{SAL02}
which homogeneously cover nearly 279 and 15 square degrees, respectively. 
Given the level of uncertainty, the design of the models is
deliberately simple. Typically they account for
two double-exponential disks (a thin and a thick one), a spheroidal halo
and a central bulge. 
\footnote{Note that the integration in equation (\ref{gflux}) can be
performed using standard numerical techniques with no need
to use slowly converging Monte Carlo methods as in Ref. \cite{BAL98}.}
Given the degree of approximation of our calculations,
we neglect the spiral structure in the disk.
The corresponding best-fitting parameters are listed in Table \ref{gal_table}.
The results are strikingly similar but for the scale height of the thick
disk, which, however, is poorly determined in both cases. 
Note that
these smooth models are far from being perfect, and neglect many detailed
features of the stellar distribution in the Galaxy.
For instance, the outer halo might not be smooth but 
be composed by overlapping streams of stars (e.g. \cite{IAL02}).
In general, residuals show that these models
systematically overpredict stellar densities towards
the galactic centre and underpredict them in the outer galaxy
\cite{SAL02}.

Because of the lack of low-latitude data, these models constrain
the inner structure of the Galaxy only weakly.
For instance, Siegel et al. \cite{SAL02} assume an $r^{-3}$ density
distribution for the bulge, while 
it has long been known that the central region of the Galaxy has a distinct
stellar density profile that approximately follows an $r^{-2}$ law
(e.g. \cite{BN68}).
Recently, it has become widely accepted that our Galaxy is barred, as evidence
accumulated over the last few years from a number of different observations.
Along this line, 
a more accurate model for the light distribution in the inner Galaxy has been 
extracted 
by Binney, Gerhardt \& Spergel \cite{BGS97} from the analysis 
of dust-corrected maps of the near-infrared brightness 
emission of the Galaxy (see e.g. \cite{DAL95}).
The corresponding parameters are listed in Table \ref{gal_table}.
However, converting near-IR brightness maps into stellar
densities its not an easy task since one has first to assume a mass-to-light 
ratio for the different components and then a stellar mass function. 
All these steps have a rather large error, and this leads to a large 
uncertainty in the final result.
On the other hand, one could think about relating IR emission directly to
either stellar density or, even, to neutrino luminosity. However, 
one needs to remember that a significant 
fraction of the IR brightness comes from dust and gas. Moreover, 
during the evolution out of the MS, the neutrino emission rate does not
correlate anymore with the photon luminosity of a star, since the only
source of thermo-nuclear neutrinos is a H burning
shell, while the structure is partially supported by the triple $\alpha$
reactions. 
For simplicity, we assumed a constant conversion factor between 
infrared brightness and stellar number density in the different components.
In the end, in order to bracket our predictions for the Galactic neutrino
background, we combine the data listed in Table \ref{gal_table} into two
distinct models.
The first one (G1) is completely based on the results by Siegel et al. 
\cite{SAL02} and it is expected to give the highest neutrino flux.
The second model (G2) is instead obtained by combining the disk+bulge fit by 
Binney et al. \cite{BGS97} with the halo+thick disk results by Chen et al. 
\cite{CAL01}.
In both cases,  whenever a given parameter is associated with
a finite range of values in Table \ref{gal_table}, we assumed the mean value 
of the quoted extremals.
Both models are normalized by imposing that the number density of stars
in the solar neighborhood belonging to the thin disk
is 0.1 pc$^{-3}$ \cite{JW97}. 
Moreover, in order to avoid unphysical divergences,  
we adopt a constant density distribution within the inner
20 pc of the bulge (in G1) and within the inner 240 pc of the halo (in both
G1 and G2).

\begin{deluxetable}{lccc}
\tabletypesize{\scriptsize}
\tablecaption{Structure parameters in the adopted Galaxy models \label{gal_table}}
\tablewidth{0pt}
\tablehead{
\colhead{Parameter} & \colhead{Binney et al. (1997)}   & \colhead{Chen et al.
(2001)}
&\colhead{Siegel et al. (2002)}}
\startdata
Disk density law & Double exponential & Double exponential& Double exponential
\\
Disk scale length (pc) & 2500 & 2250 & 2000-2500 \\
Disk scale height (pc) & 210 + 42\tablenotemark{a}&  330 & 350 \\
Disk local normalization (pc$^{-3}$) & 0.1 & 0.1 & 0.1 \\
\\
Thick-disk density law & & Double exponential & Double exponential\\
Thick-disk scale length (pc) & & 3500 & 3000-4000 \\
Thick-disk scale height (pc) & & 580-750& 900-1200 \\
Thick-disk/disk local & & & \\
\ \ \ normalization (\%)& &6.5-13 & 6-10 \\
\\
Halo density law & & Power law & Power law\\
Halo power-law index  & & 2.5 & 2.75 \\
Halo axial ratios $(b/a,c/a)$ & & (1,0.55) & (1,0.5-0.7)\\
Halo/disk local normalization (\%) & & 0.125 & 0.15\\
\\
Bar/Bulge density law & Exp-truncated power-law & & Power-law\tablenotemark{b}
\\
Bar/Bulge axial ratios $(b/a, c/a)$ & $(0.5, 0.6)$ & & $(1,0.5)$\\
Bar/Bulge power-law index & 1.8 & & 3\\
Bar/Bulge scale length (pc)& 1900 & & \\
Bar/Bulge core size (pc)& 100 & & \\
Angle between the Bar/Bulge & 
$20^\circ$ & &  \\
\ \ \ major axis and the Sun-centre line & & & \\
Bulge/disk local normalization (\%) & $6.1\times 10^{-13}$& & 0.02 \\
\\
Solar distance & & & \\
\ \ \ from the Galactic centre (pc) & 8000 & 8000 & 8000 \\
Solar distance & & & \\
\ \ \ from the Galactic mid-plane (pc) & 14 & 27 & 15
\enddata
\tablenotetext{a}{In this model the vertical structure of the disk
is given by the linear combination of two exponential laws with
different scale heights. See Binney et al. (1997) for details.}
\tablenotetext{b}{The bulge structure has been fixed a priori
in this model and it is not determined by the data}
\end{deluxetable}

No compelling constraints currently exist on the star formation histories
of the various galactic components. We adopt an overall picture based on the
studies by Binney, Dehnen, \& Bertelli \cite{BDB00} and Liu \& Chaboyer
\cite{LC00}.
In particular, we assume that the thin disk started forming
11.2 Gyr ago (which correspond to $z\simeq 2.1$ in our
cosmological model) and kept forming stars at the same pace till to the
present epoch. This is consistent with present data and with the
assumption of a Salpeter IMF \cite{BDB00}. Note, however, that 
the slope of the initial mass function near $1\,M_\odot$ 
proves to be degenerate with the rate at which the star formation rate 
declines. 
We assume a synchronous formation of the galactic halo which dates back to
12.2 Gyr ago ($z\simeq 3$), and a constant SFR in the thick disk lasting from
12.2 to 11.2 Gyr. 
As regards the bulge,
recent HST/WFPC2 \cite{FG00} and near-IR data 
\cite{ZAL03} seem to suggest that it is coeval with the halo 
and that no significative amount of new stars formed after
the intense initial starburst activity.

\begin{figure}
\centerline{\includegraphics[width=7.5cm]{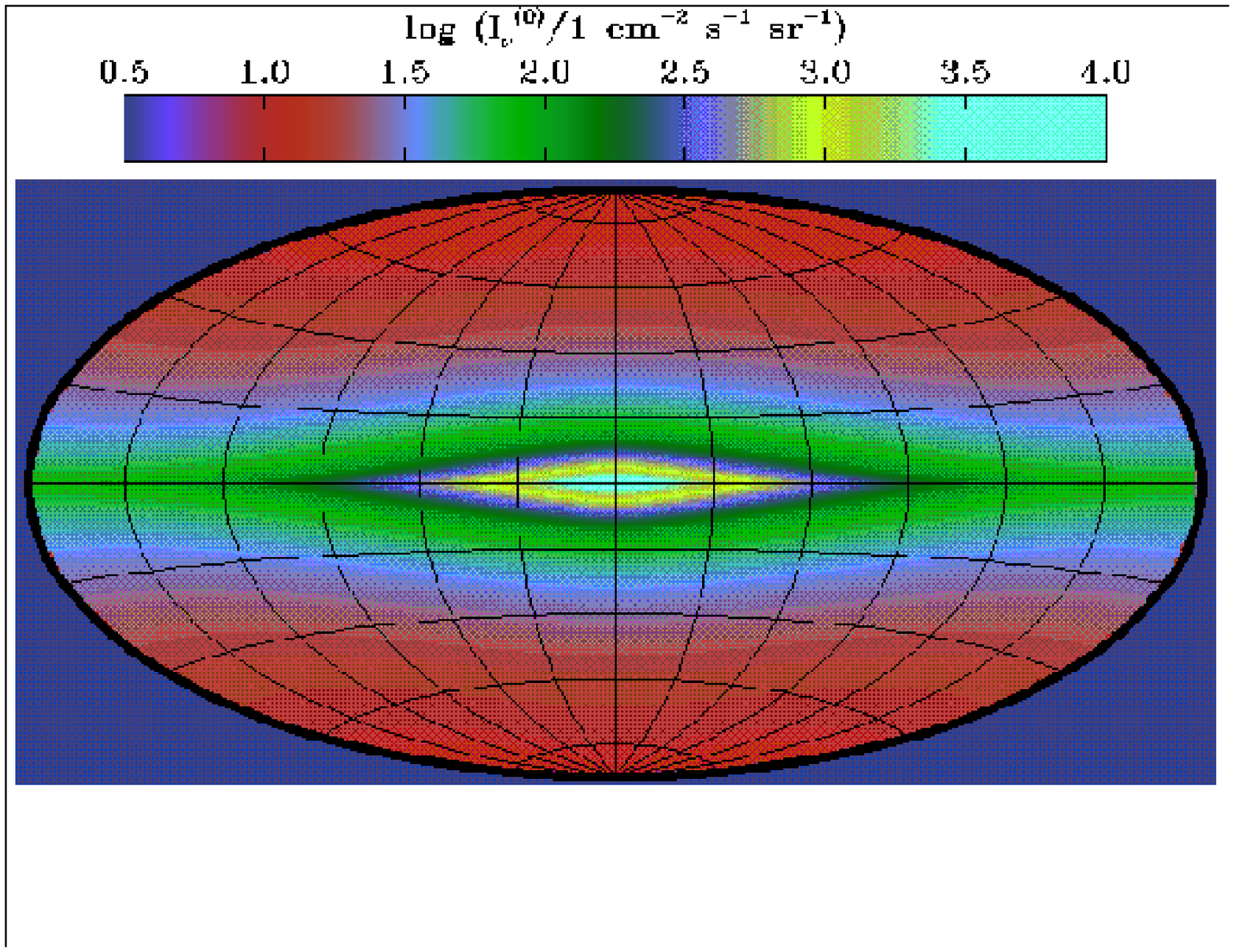}
\includegraphics[width=7.5cm]{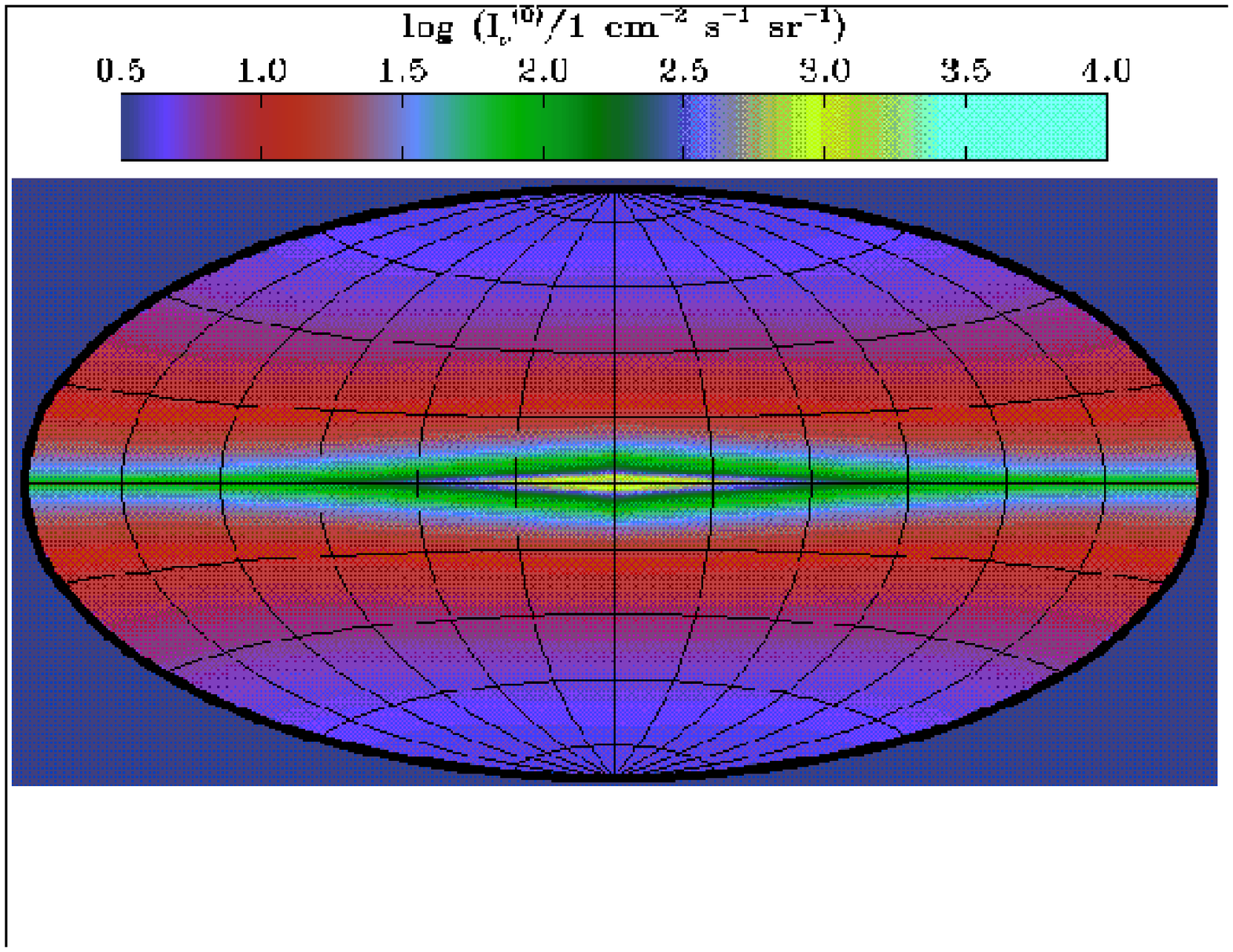}}
\caption{Aitoff projection of the intensity of the Galactic neutrino 
background on the sky.
Color coding scales logarithmically with the background intensity,
and Galactic coordinates are used.
Left and right panels refer to model G1 and G2, respectively.}
\label{maps}
\end{figure}
\begin{figure}[t]
\centerline{\includegraphics[height=8.0cm]{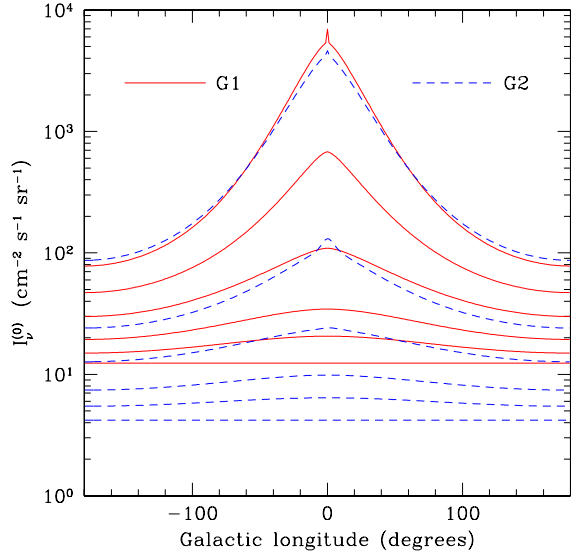}}
\caption{The Galactic neutrino background intensity as a function of direction
of observation on the sky. Different lines show the $\ell$ dependence
at fixed latitude $b$. From top to bottom, $b=0^\circ,6^\circ,15^\circ,
30^\circ,45^\circ,90^\circ$.
Continuous lines refer to model G1, while dashed lines are for G2.}
\label{phi}
\end{figure}
Sky-maps of the Galactic neutrino background intensity are shown in Figure 
\ref{maps}.
To facilitate a quantitative reading of the plots,
in Figure \ref{phi}, the background intensity is plotted
as a function of Galactic longitude for a number of 1D-cuts at
fixed latitude.
\footnote{Note that our maps have infinite angular resolution and should be 
convolved with an instrumental window function in realistic conditions.}
Both figures have been obtained by integrating $I_\nu^{(0)}(\hat{\bf r},E)$
over the particle energy, keeping $\hat{\bf r}$ fixed.
Note the spike in the background intensity
towards the galactic centre, caused by the extreme stellar
densities in the bulge.
Independently of the Galaxy model, the amplitude of the neutrino 
background varies by roughly 3
orders of magnitude between the Galactic centre and the high-latitude
isotropic component.
Because of this large variation, all the different structural component of the 
Galaxy are clearly discernible from the maps.
The amplitude of the high-latitude Galactic signal is always 
larger than the cosmic background (by a factor of a few
in G1 and by roughly an order of magnitude in G2).
This makes the detection of the cosmic background extremely difficult.

\begin{figure}[t]
\centerline{\includegraphics[height=8.0cm]{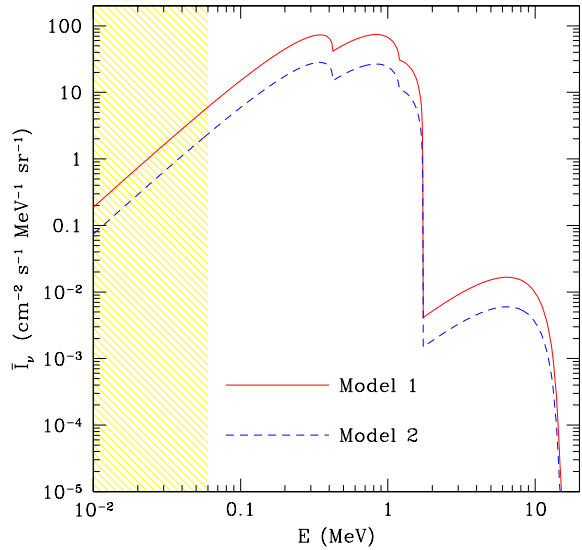}}
\caption{Energy spectrum of the Galactic neutrino background averaged
over the whole sky.}
\label{gspec}
\end{figure}

Integrating the background over the whole sky, one obtains
a neutrino flux of 943 ${\rm cm}^{-2}\,{\rm s}^{-1}$, for model 1, and
333 ${\rm cm}^{-2}\,{\rm s}^{-1}$, for model 2.
The corresponding energy spectra are shown in Figure \ref{gspec}.
In Table \ref{tbl-3} we list the contributions to the flux 
from the different structural components of the Galaxy.
At every latitude,
the counts are  dominated by neutrinos emitted in the Galactic disk
which, independently of the Galaxy model considered, account for more than
95 per cent of the total.

\begin{deluxetable}{cccccc}
\tabletypesize{\scriptsize}
\tablecaption{Galactic neutrino background. \label{tbl-3}}
\tablewidth{0pt}
\tablehead{
\colhead{Galaxy model} &
\colhead{Total} & \colhead{Thin disk}   & \colhead{Thick disk}
&\colhead{Halo}&
\colhead{Bulge}
\\
& \colhead{$({\rm cm}^{-2}\,{\rm s}^{-1})$}
& \colhead{$({\rm cm}^{-2}\,{\rm s}^{-1})$}
& \colhead{$({\rm cm}^{-2}\,{\rm s}^{-1})$}
& \colhead{$({\rm cm}^{-2}\,{\rm s}^{-1})$}
& \colhead{$({\rm cm}^{-2}\,{\rm s}^{-1})$}}
\startdata
1 & 943 & 923 & 17.4 & 1.72 & 0.66 \\
\\
2 & 333 & 314 & 15.3 & 1.12 & 3.17 \\
\enddata

\end{deluxetable}

\section{Concluding Remarks}
We have estimated two neutrino backgrounds on the sky, in the 
50 keV-10 MeV window: the Galactic component from stars presently 
burning Hydrogen and Helium in the Milky Way and
the cosmic contribution from thermonuclear reactions 
in other galaxies. 
On the grounds of the recent findings 
of Super-Kamiokande, SNO, and KamLAND, these neutrinos, 
born as $\nu_e$, propagate as a superposition of (at least) two mass 
eigenstates. When (and if) interacting with matter, they will show up as 
neutrinos of all flavors, with comparable probabilities.

Depending only slightly on some model details 
(namely, the assumed functional form of the SFR and the IMF),
the background intensity of the cosmic component peaks
at $E\simeq 0.1-0.3$ MeV, where $\bar{I}^{(0)}_\nu=3-7\,
{\rm cm}^{-2}\,{\rm s}^{-1}\,{\rm MeV}^{-1}\,{\rm sr}^{-1}$, 
and sharply drops by 2 orders of magnitude above 1-2 MeV.
The corresponding flux at the Earth's surface (integrated
over all energies) ranges between 20 and 65 ${\rm cm}^{-2}\,{\rm s}^{-1}$,
while the associated energy density is 
$\simeq 5\times 10^{-4} \,{\rm eV}\,{\rm cm}^{-3}$.
Assuming that the sum of the masses of the three neutrino
flavors is at least 1/30 eV (as suggested by 
the oscillation patterns of solar and atmospheric neutrinos) implies
that
this energy density is negligibly small with respect to the contribution
from the thermal ($T_{\rm rel}=1.95$ K)  background of relic neutrinos 
originated in
the early universe  ($u_{\rm rel} > 3.7\, {\rm eV}\,{\rm cm}^{-3}$).
Similarly, the number and entropy
densities of cosmic thermonuclear neutrinos ($n\simeq 
10^{-9}\,{\rm cm}^{-3}$, $s\simeq 10^{-7}\,k_{\rm b}\,{\rm cm}^{-3}$) 
are overwhelmed by the corresponding quantities for the primordial 
background ($n_{\rm rel} \simeq 10^2\,{\rm cm}^{-3}$,
$s_{\rm rel}\simeq 10^{3}\,k_{\rm b}\,{\rm cm}^{-3}$).

Thermonuclear neutrinos reaching the Earth today have been typically
produced by stars around $z\sim 2$. Even though they have
been around for a significant fraction of the life of the universe, 
they hardly interacted with anything. 
As far as we know, the thermonuclear neutrino background is absolutely 
transparent to 
any diffuse radiation impinging onto it since the mean free path $\lambda 
=(n\, \sigma)^{-1}$ exceeds the Hubble length for any reasonable value for 
the interaction cross section $\sigma$.

Attributing a fraction $f$ of the observed 
$\gamma$-ray background in the sub-MeV region
to neutrino radiative decay, one can 
derive a lower bound for the radiative lifetime, valid 
for any neutrino type and mass $m$, $\tau_\gamma/m \geq (2-6)\,f^{-1} 
\times 10^{12}$ s ${\rm eV}^{-1}$. This can be easily translated
into a limit for the electromagnetic form factors, which, after accounting
for the contribution of active galactic nuclei 
(i.e. by assuming $f\sim 10^{-2}$), 
is competitive with the results obtained from the 
analysis of the SN1987A neutrino burst.

Independently of the direction of observation on the sky,
the Galactic thermonuclear background dominates its cosmic counterpart.
Even at high latitudes, 
Galactic neutrinos outnumber the cosmological background
by a factor which ranges between a few and 10 depending on 
the details of the Galaxy model (see Figures \ref{maps} and \ref{phi}).
The integrated flux of Galactic neutrinos 
over the whole sky ranges between 300 and 1000
${\rm cm}^{-2}\,{\rm s}^{-1}$.
%
The emission from stars in the Galactic disk contributes more than 95
per cent of the signal.

Is there any prospect for 
detecting the Galactic component and, in the long term, to derive 
an observational neutrino map of the Galaxy? This clearly requires 
directional detectors, since one has first to distinguish the Galactic 
neutrino flux from the solar component. 
In the last ten years, large Cherenkov detectors have been built and 
successfully operated (Kamiokande, Super-Kamiokande and SNO), 
recording the directional signal of target electrons scattered from the 
impinging neutrinos. 
Analyzing
the angular distribution of the events with respect to the Sun direction, 
the peak due to solar neutrinos 
is clearly discernible over a flat background \cite{FAL01}. 
So far, this method has been applied to the 
detection of the most energetic solar neutrinos originating from Boron decay
(corresponding to a measured flux of $2.4\times 10^{6}\,{\rm cm}^{-2}\,
{\rm s}^{-1}$).
In this case,
the decay of $^{222}$Rn in water, natural radioactivity, reactor antineutrinos,
and radioactive decay of muon induced 
spallation products account for most of the instrumental background. 
After applying 
all possible cuts, background counts above 5 MeV are about ten times the 
number of events from solar neutrinos.
Even though this is not the optimal energy window for the detection of 
Galactic neutrinos, it is interesting to note that,
for $E_\nu>5$ MeV, their flux amounts to $\sim 1$~cm$^{-2}$ s$^{-1}$, which
is nearly 7 orders of magnitude smaller than the instrumental background.
This shows how far we are from their detection.

Dedicated experimental designs and techniques should then be devised to
detect cosmic and Galactic neutrinos.   
Let us consider, for instance, the cosmic (anti)neutrino background from 
Type II supernovae.
Kaplinghat et al. \cite{KSW00} found an upper
bound to the flux of
54~cm$^{-2}$ s$^{-1}$ while, more recently, Ando, Sato, \& Totani \cite{AST03}
estimate $c\cdot n\simeq 11$~cm$^{-2}$ s$^{-1}$.
Even though there is no energy region in which the contribution
from supernova neutrinos dominates over the backgrounds,
their presence might be detectable with Cherenkov telescopes 
as a distortion in the spectrum of electron and positrons produced in
the decay of invisible muons.
Ten years of integration at SuperKamiokande 
(or one at the proposed HyperKamiokande) should already guarantee 
a $1\, \sigma$ detection of the supernova neutrino background 
at $\sim 20$ MeV \cite{AST03}.

Over the next decade, the focus of solar neutrino experiments will shift
gradually from the relatively high energy $^{\mathrm 8}$B neutrinos to
the low energy, $^{\mathrm 7}$Be, p-p, and CNO neutrinos.
This is an encouraging perspective for the detection of a Galactic signal 
since, for $E<15$ MeV, the number density of thermonuclear 
neutrinos 
in the cosmic and Galactic backgrounds is expected to exceed that generated 
by Type II supernovae.
Regrettably, instrumental backgrounds tend to increase in the MeV region.
Therefore, in order to use the solar neutrino experiments to measure 
a Galactic 
signal, it is mandatory to suppress the noise in the detectors by
several orders of magnitude by using more efficient screening procedures 
and accurate selection of radiopure materials.
Note that extremely long times of integration might be anyway necessary to 
detect Galactic neutrinos with high statistical significance.

In summary, thermonuclear reactions in the Galaxy produce an anisotropic 
neutrino background in the MeV region. At the Earth's surface, this 
corresponds to an integrated flux of $\sim1000 \ {\rm cm}^{-2}\,{\rm s}^{-1}$.
Even though current detectors are unable to distinguish this signal
from other backgrounds, 
it is not inconceivable that Galactic neutrinos
might be revealed in the relatively near future with dedicated 
experimental setups.

\begin{ack}

We are grateful to Scilla Degl'Innocenti and Vittorio Castellani for
sharing the outputs of their stellar models with us, and
to John Bahcall for making the solar neutrino spectra available
in electronic format.
We thank Simon Lilly, Georg Raffelt and Mario Vietri for useful discussions.
CP has been supported by the Zwicky Prize fellowship program at ETH-Z\"urich
and he would like to express his gratitude to the Scuola Normale Superiore of 
Pisa, where this work was started, for the warm hospitality.
SP thanks the Extragalactic Astrophysics group of ETH-Z\"urich for
the kind hospitality during a visit which allowed the completion of this work.
This work has been performed within the MIUR project Astroparticle 
Physics, PRIN-2002.
\end{ack}

\appendix
\section{Entropy of a fermionic system}
Fermions have antisymmetric wavefunctions under particle exchange
and this implies that no two fermions can exist in the same 
quantum state (Pauli exclusion principle). 
The number of ways an energy level $E_i$ 
corresponding to $g_i$ degenerate quantum states can be populated
with $n_i$ fermions is given by the binomial distribution with $g_i$ boxes, 
where $n_i$ contain 1 particle and $g_i-n_i$ are empty, i.e.
\be
\frac{g_i!}{n_i!\, (g_i-n_i)!}\;.
\end{equation}
Allowing for a system with different energy levels, the total number of 
configurations is
\be
W(\{n_i\})=\prod_i \frac{g_i!}{n_i!\, (g_i-n_i)!}\;.
\end{equation}
We define the entropy of the system as 
\begin{eqnarray}
S&=&k_{\rm b} \ln W(\{n_i\})\simeq\\ 
&\simeq& k_{\rm b} \sum_i \left\{
g_i (\ln g_i-1) - n_i (\ln n_i-1) - (g_i-n_i) [\ln (g_i-n_i)-1]
\right\}\;, \nonumber
\label{entro1}
\end{eqnarray}
where, assuming both $n_i$ and $g_i$ are $\gg 1$, 
we used Stirling's approximation $\ln n!\simeq n\, (\ln n-1)$. 
The problem with eq.(\ref{entro1}) is that it requires
knowledge of the whole set $\{n_i\}$, which, in practice, 
is a formidable task. A standard approximation is to 
replace $W(\{n_i\})$ with $W(\{\bar{n}_i\})$ 
(where $\bar{n}_i$ denotes 
the mean occupation number of the energy level $E_i$) in the above definition 
of entropy. In this case, neglecting 1 compared to $g_i$, one gets
\be
S = k_{\rm b} \sum_i \left[
\bar{n}_i \ln\left(\frac{g_i}{\bar{n}_i}-1\right)
-g_i \ln\left(1-\frac{\bar{n}_i}{g_i}\right)
\right]\;.
\end{equation}
It is convenient to 
introduce the mean occupation number of a single quantum state
$f_i=\bar{n}_i/g_i$ to obtain
\be
S = k_{\rm b} \sum_i g_i \left[
f_i \ln\left(\frac{1}{f_i}-1\right)
-\ln\left(1-f_i\right)
\right]\;.
\end{equation}
If the energy levels of the system are closely spaced, one may replace the 
discrete 
probability density  $f_i$ by a continuous function $f(E)$.
This is easily done for a set of free particles moving in an (essentially)
infinite volume. In this case, all the sums over the energy levels 
$\sum_i g_i \dots$ may be replaced by integrals over the 
accessible phase-space volume
$ \int \dots d\Gamma/h^3$ with $d\Gamma=d^3q\,d^3p$.
For ultra-relativistic particles, for which $E=c\,p$, the entropy per unit 
volume is then given by:
\be
s=\frac{4\pi\,k_{\rm b}\,g_{\rm s}}{c^3\,h^3} \int_0^\infty E^2\left\{
f(E) \ln\left[\frac{1}{f(E)}-1\right]
-\ln\left[1-f(E)\right] 
\right\} dE\;,
\end{equation}
with $g_{\rm s}$ the number of helicity degrees of freedom.
This can be expressed in terms of the particle number density,
\be
n=\frac{4 \pi \,g_{\rm s}}{c^3\,h^3} \int_0^\infty E^2 \,f(E)\,dE\;,
\end{equation}
as
\be
s=\chi\,k_{\rm b}\,n\;,
\end{equation}
with
\be
\chi=\frac{\int_0^\infty E^2\left\{
f(E) \ln\left[f(E)^{-1}-1\right]
-\ln\left[1-f(E)\right] 
\right\} dE}{\int_0^\infty E^2 \,f(E)\,dE}
\end{equation}
a dimensionless coefficient.

\section{Photon Spectrum from Radiative Decays}
\label{raddec}
Let us assume that a set of cosmological sources produce neutrinos
with a rate $\dot{n}_\nu(z)$ 
(particles per unit comoving volume per unit time) 
and with a redshift dependent energy spectrum 
$S(E_\nu,z)$ (normalized to unity). 
Then, the differential comoving neutrino number density per unit energy 
at redshift $z$ is given by
\begin{eqnarray}
\label{dnu_denu}
\!\!\!\!\!\!\!\!\!\!
&\frac{dn_\nu}{dE_\nu}&(E_\nu,z)=\int_z^\infty 
\!\!\!\dot{n}_\nu(z') 
\,S\left(\frac{E_\nu(1+z')}{1+z},z'\right)
\frac{1+z'}{1+z} \left|\frac{dt}{dz'}\right| \,dz' \\
\!\!\!\!\!\!\!\!\!\!\!\!\!\!\!\!\!\!\!\!\!\!&=&
\frac{1}{H_0\,(1+z)}
\int_z^\infty \!\!\! S\left(\frac{E_\nu(1+z')}{1+z},z'\right)
\frac{\dot{n}_\nu(z')\,dz'}
{[\Omega_0\,(1+z')^3+\Omega_k\,(1+z')^2+\Omega_\Lambda]^{1/2}}
\;.\nonumber
\end{eqnarray}

If neutrinos of energy $E_\nu$ decay radiatively with a (rest-frame)
lifetime $\tau_\gamma\gg H_0^{-1}$, the decay rate in the cosmic frame
will be $n_\nu(z)/\gamma(E_\nu)\,\tau_\gamma$,
where the Lorentz factor $\gamma(E_\nu)=E_\nu/m_\nu\, c^2$.
Assuming that the secondary neutrino is very light compared to the parent
one implies that the decay photons have $E\simeq E_\nu/2$, so that
the decay processes will produce a photon background with present-day 
intensity: 
\begin{equation}
\label{decayspec}
j(E)=\frac{c}{4\pi} \int_0^\infty \!\!\! dz \,(1+z)\int_0^\infty
\!\!\!\frac{dE_\nu}{\gamma(E_\nu)\,\tau_\gamma}
\frac{dn_\nu}{dE_\nu}(E_\nu,z)\,\delta_D\left(E(1+z)-\frac{E_\nu}{2}\right)
\left|\frac{dt}{dz}\right|\;.
\end{equation}
Inserting equation (\ref{dnu_denu}) into equation (\ref{decayspec}) and
integrating over the Dirac-delta distribution, one eventually gets
\be
\label{de}
j(E)=\frac{c}{4\pi} \,
\frac{m_\nu c^2}{\tau_\gamma}\,
\frac{1}{H_0^2\,E}\,
\int_0^\infty \!\!\! \frac{dz}{(1+z)^2\,K(1+z)}
\int_z^\infty \!\!\! dz'\,\frac{\dot{n}_\nu(z')\,S(2E(1+z'),z')}{
K(1+z')}\;,
\end{equation}
with
$K(x)=[\Omega_0\,x^3+\Omega_k\,x^2+\Omega_\Lambda]^{1/2}$.
In Section 2.7,
we used this expression to constrain the neutrino decay time.

Considering, for simplicity,
a redshift-independent, monochromatic neutrino spectrum at emission, 
$S(E_\nu,z)=\delta_D(E_\nu-E_0)$, the neutrino energy distribution reduces to
\be
\label{dnu_mono}
\frac{dn_\nu}{dE_\nu}(E_\nu,z)=
\begin{cases}
\displaystyle{
\frac{E_\nu^{1/2}\,\dot{n}_\nu([E_0(1+z)/E_\nu]-1)} {H_0\, 
[\Omega_0\,E_0^3\,(1+z)^3+\Omega_k\,E_\nu\,E_0^2\,(1+z)^2+\Omega_\Lambda\,
E_\nu^3]^{1/2}}}
&\text{for} \quad E_\nu\leq E_0\\
0 &\text{for} \quad E_\nu>E_0
\end{cases} 
\end{equation}
where cosmological expansion smeared the energy of neutrinos into a continuous
spectrum. Similarly, for the photon background intensity, one gets
\be
j(E)=\begin{cases}
\displaystyle{
\frac{c}{4\pi}\,
\frac{1}{2}\,
\frac{m_\nu c^2}{\tau_\gamma}\,
\frac{\,\dot{n}_\nu(\tilde{z}(E))}
{H_0^2\,E^2\,
K(E_0/2E)}
\,\int_1^{E_0/2E}
\!\!\!\!\!\!\!\frac{dw}{w^2\,K(w)}}
&\text{for} \quad E_\nu\leq E_0/2\\
0 &\text{for} \quad E_\nu>E_0/2
\end{cases} 
\end{equation}
with $\tilde{z}(E)=(E_0/2E)-1$.
For an Einstein-de Sitter universe ($\Omega_0=1, \Omega_\Lambda=0$), 
this becomes
\be
\label{eds}
j(E)=
\begin{cases}
\displaystyle{
\frac{c}{4\pi}\,
\frac{2^{3/2}}{5}
\frac{m_\nu c^2}{\tau_\gamma}\,
\frac{\dot{n}_\nu(\tilde{z}(E))}
{H_0^2\,E_0^2}\,
\left(\frac{E_0}{E}\right)^{1/2}\,\left(1-\left(\frac{2 E}{E_0} \right)^{5/2}
\right)}
&\text{for} \quad E\leq E_0/2\\
0 &\text{for} \quad E>E_0/2
\end{cases} 
\end{equation}
with $j(E)\propto E^{-1/2}$ for $E\ll E_0$.
\footnote{
Note that the photon spectra published in Ref. \cite{KT89}
miss the last factor on the right hand side of equation (\ref{eds}).
In consequence, they are accurate only for $E \ll E_0$,
which corresponds to extremely high-redshift sources of neutrinos which are
not relevant for our calculations.}
On the other hand, an empty universe ($\Omega_0=\Omega_\Lambda=0$) 
would give
\be
j(E)=
\begin{cases}
\displaystyle{
\frac{c}{4\pi}\,
\frac{1}{2}\,
\frac{m_\nu c^2}{\tau_\gamma}\,
\frac{\dot{n}_\nu(\tilde{z}(E))}
{H_0^2\,E_0^2}\,
\left(\frac{E_0}{E}\right)\,\left(1-\left(\frac{2 E}{E_0} \right)^{2}
\right)}
&\text{for} \quad E\leq E_0/2\\
0 &\text{for} \quad E>E_0/2
\end{cases} 
\end{equation}
Finally, the effect of a non-vanishing cosmological constant can be sketched
by looking at a de Sitter universe ($\Omega_0=0, \Omega_\Lambda=1$)
\be
j(E)=
\begin{cases}
\displaystyle{
\frac{c}{4\pi}\,
\frac{1}{2}
\frac{m_\nu c^2}{\tau_\gamma}\,
\frac{\dot{n}_\nu(\tilde{z}(E))}
{H_0^2\,E_0^2}\,\left(\frac{E_0}{E}\right)^2\,\left(1-\left(\frac{2 E}{E_0} \right)
\right)}
&\text{for} \quad E\leq E_0/2\\
0 &\text{for} \quad E>E_0/2\;.
\end{cases} 
\end{equation}

\end{document}